\documentclass{iopart}
\usepackage{times}
\usepackage{color}

\newif\ifpdf
\ifx\pdfoutput\undefined
\pdffalse % we are not running PDFLaTeX
\else
\pdfoutput=1 % we are running PDFLaTeX
\pdftrue
\fi

\ifpdf
\usepackage[pdftex]{graphicx}
\else
\usepackage{graphicx}
\fi

\newcommand{\Label}[1]{\label{#1}}
\newcommand{\BEC}{Bose-Einstein condensate}

\def\DRAFT{%
\renewcommand{\Label}[1]{\label{##1}%\message{##1}%
{\hbox to 0cm{\textcolor{green}{\hss\em ##1\quad}}}}}
\newcommand{\HFB}{{Hartree-Fock-Bogoliubov}}
\newcommand{\ccfs}{cold collision frequency shift} 

\def\FIGURE#1#2{
\begin{figure}
{\hskip25mm\includegraphics[width=10.6cm]{#1}}
\caption{\label{#1}#2}
\end{figure}}

\begin{document}
%%%%%%%%%%%%%%%%%%%%%%%%%%%%%%%%%%%%%%%%%%%%%%%%%%%%%%%%%%%%%%%%%%%%%%%%%%
\ifpdf
\DeclareGraphicsExtensions{.pdf, .jpg}
\else
\DeclareGraphicsExtensions{.eps, .jpg}
\fi
%%%%%%%%%%%%%%%%%%%%%%%%%%%%%%%%%%%%%%%%%%%%%%%%%%%%%%%%%%%%%%%%%%%%%%%%%%
\bibliographystyle{iopacm}
\title[Theory of \ccfs\ in hydrogen]{Theory of the \ccfs\ in 1S--2S 
spectroscopy of  
Bose-Einstein-condensed and non-condensed hydrogen}
\author{C.W. Gardiner and A.S. Bradley}
\address{School of Chemical 
and Physical Sciences, Victoria University, Wellington, New Zealand} 
%\maketitle
\begin{abstract}
We show that a correct formulation of the \ccfs\ for two photon 
spectroscopy of Bose-condensed and cold non-Bose-condensed hydrogen is 
consistent with experimental data.  Our treatment includes transport 
and inhomogeneity into the theory of a non-condensed gas, which causes 
substantial changes in the \ccfs\ for the ordinary thermal gas, as a 
result of the very high frequency (3.9kHz) of transverse trap mode.  
For the condensed gas, we find substantial corrections arise from the 
inclusion of quasiparticles, whose number is very large because of the 
very low frequency (10.2Hz) of the longitudinal trap mode.  These two 
effects together account for the apparent absence of a ``factor of 
two'' between the two possibilities.

Our treatment considers only 
the Doppler-free measurements, but could be extended to 
Doppler-sensitive measurements.  For Bose-condensed hydrogen, we 
predict a characteristic ``foot'' extending into higher detunings 
than can arise from the condensate alone, as a result of a correct 
treatment of the statistics of thermal quasiparticles.

\end{abstract}
%\StartTwoColumn

\section{Introduction}
Bose-Einstein condensation of spin-polarized hydrogen, achieved in 
1998 
\cite{Fried1998a,Killian1998a,Killian2000a,KillianThesis,FriedThesis} 
as the culmination of a quarter of a century of development of 
cooling, trapping and spectroscopic techniques, still presents certain 
puzzling features.  The most notable of these occurs in the 
spectroscopic method used to detect the condensate, based on 
two-photon excitation from the 1S to the 2S state of the hydrogen atom 
in the presence of a significant background density of hydrogen 
atoms---a density which of course becomes very large when the hydrogen 
Bose condenses.  The {\em cold collision frequency shift}---a shift of 
the transition frequency proportional to the density of the gas---is 
measurable even when the gas is merely dense, but not Bose-condensed, 
and is very significant for the condensate.  This has led to the 
use---suggested by Shlyapnikov (see reference [19] of 
\cite{Cesar1996a})---of the cold 
collision frequency shift as a method of measuring the proportion of 
atoms which occur at a given density in the system of cold hydrogen 
atoms.  The magnitude of the shift is expected to be twice as much for 
a given density for the non-condensed gas as for the Bose-condensate, 
as a result of the different coherence properties of the two systems.  
However, the frequency shifts observed have led the experimenters to 
the conclusion \cite{Fried1998a}:

\begin{quote}
\dots if we had assumed that the frequency shift in
the condensate is only half as large as for a normal gas
at the same density, as one would expect for a condensate
in a single quantum state, the density extracted from the
spectrum would be twice as large. This would imply that
$ N_c=6\times10^9 $ and yield an unreasonably high condensate
fraction of 25\%.
\end{quote}

The apparent absence of the ``factor of two'' has led to the 
speculation that the hydrogen condensate is not in fact a true 
condensate, but is in some unexplained way incoherent, so that the 
behaviour is qualitatively similar to that of a non-condensed system, 
and C\^ot\'e and Kharchenko \cite{cote1999ax} have given a more exotic 
explanation, 
based on droplet formation at the edge of the condensate.
Our aim in this paper is to take the view that the hydrogen \BEC\
has no exotic features, but to make a careful investigation of the 
physics of the two-photon  excitation process used to probe the 
system, both when it is condensed and when it is non-condensed.  In 
our view, the apparent absence of the ``factor of two''
 implies that something is wrong with the theoretical treatment of 
 the \ccfs\ in {\em either} the condensed system {\em or}
 the non-condensed system or possibly in both, as they have been so 
 far formulated.
 
 We have therefore reviewed the conventional theory in Sect.1, and 
 have shown that for a non-condensed system it is assumed that the 
 system is homogeneous.  The treatment of the condensed system
 \cite{KillianThesis,Killian2000a} does not
make this assumption, but requires the system to be a pure 
condensate.  Neither of these requirements can be considered to be 
adequately met in current experiments, since the system is 
tightly trapped and relatively warm.

In Sect.2 we develop a theory of the \ccfs\ in inhomogeneous 
non-condensed systems, and show that in the published experiments
\cite{Killian1998a,Fried1998a} the effect of the inhomogeneity will 
be very significant, and will tend to reduce the ``factor of two'' 
significantly.  The main reason is quite straightforward.  During the 
time of the exciting pulse the atoms move through a wide range of 
densities, and the coherence between the excited state and the ground 
state, necessary to produce the ``factor of two'', cannot be 
established.

In Sect.3 we turn to the condensed system, and develop a formalism 
which uses a quasiparticle basis to give some linear partial 
differential equations for the excitation as a function of space and 
time.  We use the very anisotropic aspect ratio of the hydrogen 
trap---about 400:1---to simplify our formalism to a relatively simple 
set of formulas from which quantitative predictions can be made. Our 
results confirm those of Killian \cite{Killian2000a} for a pure 
condensate, but also give results which are valid when a significant 
quasiparticle component is present.  In essence we show that the 
quasiparticle component can be up to 20\%\ of the apparent condensate 
occupations, and that the condensate and the quasiparticles give rise 
to  independent and qualitatively different signals.  For a given 
excitation frequency, the two signals are produced from different 
locations in the system---this can be seen as arising because there is 
not  only a \ccfs\, but as well a splitting which gives rise to two 
frequencies, the two components corresponding to the two signals.  The 
situation is illustrated in Fig.\ref{figQC}.  In fact, at the 
highest detunings, the signal arises only from the quasiparticle 
component, and gives rise to a characteristic ``foot'' protruding 
into the high frequencies above an otherwise smooth spectrum, and 
indeed there is some experimental indication of such a feature---see 
Fig.\ref{fig6}.

We conclude that if the our calculations for condensed and 
non-condensed experiments are taken into account that the data 
of \cite{Fried1998a}
appear to be consistent with the following:
\begin{itemize}
\item[i)] The calculated value of the 1S-2S cross section.
\item[ii)] The condensate is coherent, and is accompanied by the 
appropriate distribution of quasiparticles for the measured 
temperature.
\item[iii)] The non-condensed system is incoherent, with 
$g_2=2$.

\end{itemize}

%%%%%%%%%%%%%%%%%%%%%%%%%%%%%%%%%%%%%%%%%%%%%%%%%%%%%%%%%%%%%%%%%%%%%%%%%%%%

\section{The cold collision frequency shift for a homogeneous system}
In an inhomogeneous and possibly Bose condensed system, we consider the 
excitation by a coherent process. One photon excitation by a coherent 
microwave  
field is usual for an atomic clock, but in the case of a hydrogen \BEC\    
two-photon excitation is used to excite from the 1S to the 2S level.  The 
exciting field is pulsed for a time very long compared to the transition 
frequency, 
but so short that only an infinitesimal proportion of the atoms are excited to 
the upper level.
\subsection{Second-quantized Hamiltonian}
For the case under consideration the 
second-quantized Hamiltonian can be written in the form
\begin{eqnarray}\fl\Label{a1}
H &=& \int d^3{\bf x}\, \Bigg\{ 
\psi^\dagger_1\left(T + V({\bf x}) +\hbar\omega_1\right)\psi_1
+
\psi^\dagger_2\left(T + V({\bf x}) +\hbar\omega_2\right)\psi_2
\nonumber \\ \fl
&& + {u\over 2}\psi^\dagger_1\psi^\dagger_1\psi_1\psi_1
 + {v}\psi^\dagger_1\psi^\dagger_2\psi_1\psi_2
 + {w\over 2}\psi^\dagger_2\psi^\dagger_2\psi_2\psi_2
+ gE({\bf x})\bigg(
\psi^\dagger_1\psi_2 e^{i\Omega t} + \psi^\dagger_2\psi_1 e^{-i\Omega t}
\bigg)\Bigg\}
.
\nonumber\\ \fl
\end{eqnarray}
\begin{itemize}
\item[i)]
The labels $ 1,2$ represent the 1S and 2S states of hydrogen, and 
$\hbar\omega_1, \hbar\omega_2  $ are 1S and 2S energy levels used in the 
experiment.
\item[ii)]
The kinetic energy operator is $ T = -\hbar^2\nabla^2/2m$
\item[iii)]
This form assumes the trapping potentials  have the same form
$ V({\bf x})$ for the 1S and the 2S states of hydrogen.
\item [iv)]
Since the process being considered is a two-photon process, the 
driving term is proportional to the square of the laser field, which 
is assumed to have frequency $\Omega/4\pi$, and wavenumber 
${\bf k}/2$, so that we can write the driving term using
\begin{eqnarray}
E({\bf x}) &=& E\big(1 + \cos({\bf k}\cdot{\bf x})\big).
\end{eqnarray}
The spatially independent term gives the {\em Doppler-free} 
excitation, while the cosine term gives the {\em Doppler-sensitive} 
excitation.  In the experiments both forms of excitation are 
used---however in this paper we shall treat mainly the Doppler-free case, 
for which we have been able to develop a relatively simple formalism.

\item[v)]
The interaction coefficients are given in terms of scattering lengths by
\begin{eqnarray}\Label{a101}
u &=& {4\pi a_{\rm 1S-1S}\hbar^2/m}
\\ \Label{a10101}
v &=& {4\pi a_{\rm 1S-2S}\hbar^2/m}
\\ \Label{a10102}
w &=& {4\pi a_{\rm 2S-2S}\hbar^2/m}
\end{eqnarray}
The scattering lengths have been calculated in 
\cite{Jamieson1995a,Jamieson1996a} to be
\begin{eqnarray}\Label{a102}
a_{\rm 1S-1S} &=& \phantom{-}0.0648{\rm nm} 
\\ \Label{a10201a}
a_{\rm 1S-2S} &=& -2.3{\rm nm},
\end{eqnarray}
and the experimental value  (calculated assuming that the theory of 
the \ccfs\ in a non-condensed trapped system is the same as that of a 
homogeneous trapped system)
\begin{eqnarray}\Label{a10201b}
a_{\rm 1S-2S} &=& -1.4\pm 0.3{\rm nm}.
\end{eqnarray}
is reported in \cite{Killian1998a}.  

\item[vi)]
The coefficient $ w$ will be set equal to zero, since its effect is in 
practice negligible with the densities of 2S atoms appropriate for the 
experiment.
\end{itemize}
\subsection{Background}\label{background}
The approach used by Killian \cite{Killian2000a,KillianThesis} considers an 
initial many-body wavefunction composed as a symmetrized product of one-body 
wavefunctions $  | 1S \rangle $.  The result of the laser  excitation 
is to rotate the spin wavefunction slightly into the 2S subspace, so the 
one-body wavefunction is transformed
\begin{eqnarray}\Label{bg1}
 | 1S \rangle \to  t | 1S\rangle+ r | 2S\rangle ,
\end{eqnarray}
with $ |t|^2 + |r|^2 =1$, and $ r \ll 1$.  In field theoretic language (in the 
Heisenberg picture) the initial state is described by field operators 
$ \psi_1({\bf x})$, $ \psi_2({\bf x})$ with 
$ \langle \psi^\dagger_2({\bf x})\psi_2({\bf x})\rangle =0$, and these
transform to
\begin{eqnarray}\Label{bg2}
\psi_1({\bf x}) & \to & \phantom{-}t \psi_1({\bf x}) + r \psi_2({\bf x})
\\ \Label{bg3}
\psi_2({\bf x}) & \to & -r \psi_1({\bf x}) + t \psi_2({\bf x})  .
\end{eqnarray}
and the population densities 
$ n_i({\bf x}) =  \langle \psi^\dagger_i({\bf x})\psi_i({\bf x})\rangle$
transform as
\begin{eqnarray}\Label{bg4}
n_1({\bf x}) \to   n_1({\bf x}) - \delta n({\bf x})& =& |t|^2 n_1({\bf x}),
\\ \Label{bg5}
n_2({\bf x})  \to   
 \delta n({\bf x})\,\,\,\,\qquad\quad& =& |r|^2 n_1 ({\bf x}).
\end{eqnarray}
The mean energy after this process is given by the sum of kinetic, potential 
and self energy terms. Since the trapping potentials for the two states are 
assumed the same, there is no change of trapping energy, and if the excitation 
is not strongly dependent on space there is no change in the kinetic energy 
term.  Under these conditions the main energy change is given by
the interaction energy, which  changes by an average amount
\begin{eqnarray}\fl\Label{bg6}
\left({u\over 2}\left(|t|^4-1\right) + v |t|^2|r|^2\right)
\langle\psi_1^\dagger({\bf x})\psi^\dagger_1({\bf x})
\psi_1({\bf x})\psi_1({\bf x})\rangle
& \approx&
 g_2({\bf x},{\bf x})(v-u)n_1({\bf x})\delta n({\bf x}).
\end{eqnarray}
Here we have used the spatial density correlation function 
\begin{eqnarray}\Label{bg7}
 g_2({\bf x},{\bf x}') &=&
{\langle\psi_1^\dagger({\bf x})\psi^\dagger_1({\bf x}')
\psi_1({\bf x})\psi_1({\bf x}')\rangle\over n_1({\bf x})n_1({\bf x}')}.
\end{eqnarray}
If the initial quantum state is noncondensed, it can be written in terms of a 
set of orthonormal one particle wavefunctions $ \phi_r({\bf x)}$ with  
occupations $  N_r $ which are either 0 or 1, so that
\begin{eqnarray}\Label{bg8}
n_1({\bf x}) & = & \sum_r |\phi_r({\bf x)}|^2  N_r,
\\ \Label{bg9}
g_2({\bf x}, {\bf x})n_1({\bf x})^2 &=& 
2\sum_{{r,s}\atop r\ne s}|\phi_r({\bf x)}|^2 |\phi_s({\bf x)}|^2 N_r N_s
\\ \Label{bg10}
&\approx& 2 n_1({\bf x})^2,
\end{eqnarray} 
where the  approximate final result is valid provided the occupation is spread 
over very many modes, so that the missing term with $ r=s$ is negligible.  Thus 
we find in this case $ g_2({\bf x}, {\bf x}) \to 2 $.  This result is also 
obviously true for an ensemble of noncondensed quantum states, such as a 
thermal noncondensed state.

If the state is condensed in the sense that only the quantum state $ r=r_0$ is
occupied, and $ N_{r_0}\gg 1$ , then the same calculation leads to 
$ g_2({\bf x}, {\bf x}) \to 1 $, so that the energy shift is only half of the 
value expected from a noncondensed state---the ``factor of two'' thus does 
appear when the trapping and kinetic parts of the Hamiltonian can be 
neglected.

\subsubsection{Issues for a realistic condensate}
There are three main issues which need to be addressed
\begin{itemize}
\item[i)] The energy shift is proportional to $ n_1({\bf x})$, and is 
sufficiently large in practice for it to be used as a spectroscopic method of 
measuring the density profile of a very cold and possibly condensed cloud of 
hydrogen atoms.  Thus the amplitude $ t$ for transition to the 2S state is of 
necessity spatially dependent. This issue has been addressed  in 
\cite{Killian2000a,KillianThesis} for the case of a pure condensate, 
but it has been assumed that the results of the homogeneous 
noncondensed system are applicable to the inhomogeneous non-condensed case 
which 
occurs in the experiments.  A full consideration of the effects of 
inhomogeneity for a noncondensed system or a partially condensed 
system has not been made.

\item[ii)] The temperature of the condensed system is relatively high, so that 
occupation of the lowest quasiparticle levels is significant.  We showed 
previously \cite{Gardiner0009371} that this could be up  to 20\% of the 
apparent condensate population, and that the effective $ g_2({\bf x},{\bf x})$ 
in such a 
situation is spatially dependent, varying from a little more than 1 to somewhat 
larger than 2, depending 
on the position in the condensate.  Under these conditions  the simple 
argument given above cannot be justified.

\item[iii)] The estimate of the energy shift (\ref{bg6}) uses an average 
energy.  However, at a given position in space, this could be the result of an 
average of different eigenvalues, and it would be these eigenvalues which were 
selected spectroscopically, not their average, leading to a given 
resonant frequency corresponding to different densities of the gas.

\end{itemize}
\section{Cold collision frequency shift for an inhomogeneous non-condensed 
system}

%%%%%%%%%%%%%%%%%%%%%%%%%%%%%%%%%%%%%%%%%%%%%%%%%%%%%%%%%%%%%%%%%%%%%%%%%%%%%%%
%
As noted above, the ``factor of 2'' has been derived only for 
\emph{homogeneous} systems, and in the experiments 
\cite{Killian2000a,KillianThesis} homogeneity is not satisfied.  One 
need only note:
\begin{eqnarray*}\label{}
\mbox{The radial trap frequency is }  &\approx& 3.9 \mbox{kHz}
\\
\mbox{The axial trap frequency is}  &\approx &10 \mbox{Hz}
\\
\mbox{The laser pulse length is}
&\approx& 500 \mu\rm{s}
\approx 2\mbox{ radial trap periods.}
\end{eqnarray*}
Thus, during the period of the excitation it will be possible for a 
significant number of atoms to experience a wide range of values of 
the vapour density, and a treatment must be devised which includes the 
full dynamics of transport in the tightly trapped cloud.  We will show 
that a combined transport-excitation equation in 3 position plus 3 
momentum variables can be derived.

\subsection{Equations of motion}\label{eqsmotion}
From the Hamiltonian (\ref{a1}) the Heisenberg equations of motion for 
the field operators are (in the Doppler-free case)
\begin{eqnarray}\Label{TH-d1}
i\hbar\dot \psi_1 &=& [\hbar\omega_1+H({\bf x})]\psi_1 
                     +u\psi_1^\dagger\psi_1\psi_1
                     +v\psi_2^\dagger\psi_1\psi_2  
                     +gEe^{i\Omega t}\psi_2
\\ \Label{TH-d2}
i\hbar\dot \psi_2 &=& [\hbar\omega_2+H({\bf x})]\psi_2 
                     +v\psi_1^\dagger\psi_1\psi_2
                     +gEe^{-i\Omega t}\psi_1,
\end{eqnarray}
in which
\begin{eqnarray}\Label{TH-d201}
H({\bf x}) &\equiv& -{\hbar^2\nabla^2\over 2m} +V({\bf x}).
\end{eqnarray}
We suppose that the occupation of the 1S level is always very much 
greater than that of the 2S level, and all of this occupation induced by the 
application of the coherent driving field $ E$.
%%%%%%%%%%%%%%%%%%%  NEW      !!!
This means that we can neglect entirely the last two terms in (\ref{TH-d1}), so 
that $ \psi_1({\bf x},t) $ can be regarded as a known time-dependent operator.

We now define  Wigner amplitudes  
\begin{eqnarray}\Label{TH-d3}
 f ({\bf p},{\bf x}) &=& {1\over h^3}\int d^3{\bf y}
\langle \psi_1^\dagger({\bf x}- {\bf y}/2)\psi_2({\bf x}+ {\bf y}/2)\rangle
e^{-i{\bf p}\cdot{\bf y}/\hbar}
\\ \Label{TH-d4}
 n ({\bf p},{\bf x}) &=& {1\over h^3}\int d^3{\bf y}
\langle \psi_1^\dagger({\bf x}- {\bf y}/2)\psi_1({\bf x}+ {\bf y}/2)\rangle
e^{-i{\bf p}\cdot{\bf y}/\hbar}
\end{eqnarray}
We will also use the momentum integrated quantities
\begin{eqnarray}\Label{TH-d5}
N({\bf x}) &=& \int d^3{\bf p}\, n({\bf p},{\bf x}) \equiv \
\langle \psi_1^\dagger({\bf x})\psi_1({\bf x})\rangle
\\ \Label{TH-d6}
F({\bf x}) &=& \int d^3{\bf p}\, f({\bf p},{\bf x}) \equiv \
\langle \psi_1^\dagger({\bf x})\psi_2({\bf x})\rangle
\end{eqnarray}
which will characterize the system sufficiently for our purposes.
We can now derive equations of motion using 

\begin{itemize}
\item[i)] Hartree-Fock factorization in the form (for example)
\begin{eqnarray}\fl\Label{TH-d7}
\langle \psi_1^\dagger({\bf z}) .\psi_1^\dagger({\bf z}') 
\psi_1({\bf z}')\psi_2({\bf z}')\rangle
\nonumber
\\ \fl \qquad\qquad =
\langle \psi_1^\dagger({\bf z})\psi_1({\bf z}')\rangle
\langle \psi_1^\dagger({\bf z}')\psi_2({\bf z}')\rangle
+
\langle \psi_1^\dagger({\bf z})\psi_2({\bf z}')\rangle
\langle \psi_1^\dagger({\bf z}')\psi_1({\bf z}')\rangle,   
%\nonumber\fl\\
\end{eqnarray}
with ${\bf z} \equiv {\bf x}- {\bf y}/2$ and ${\bf z}' \equiv {\bf x}+ 
{\bf y}/2$.

\item[ii)]
Since the terms 
$\langle \psi_1^\dagger({\bf z})\psi_1({\bf z}')\rangle $, 
$ \langle \psi_1^\dagger({\bf z})\psi_2({\bf z}')\rangle $
can only be nonzero for very small $ {\bf y} = {\bf z}-{\bf z}'$,
 we also make the approximations 
of the form
\begin{eqnarray}\fl\Label{TH-d8}
\langle \psi_1^\dagger({\bf z}')\psi_1({\bf z}')\rangle
&\equiv& n({\bf z}') \approx
 n({\bf x})+ {\bf y}\cdot\nabla n({\bf x}) /2 .
\end{eqnarray}
\item[iii)]  We can then use the standard Wigner function methods to get the 
equation of motion 
\begin{eqnarray}\fl\Label{TH-d9}
i\hbar {\partial f({\bf p},{\bf x}) \over\partial t } &=&
\hbar(\omega_2-\omega_1)f({\bf p},{\bf x}) + gE e^{-i\Omega t}
n({\bf p},{\bf x}) 
\nonumber\\\fl
&&+i\hbar\left\{
-({\bf p}/m)\cdot\nabla_{x} + \nabla V({\bf x})\cdot\nabla_p\right\}
f({\bf p},{\bf x})
\nonumber\\\fl
&& -(2u-v)N({\bf x})f({\bf p},{\bf x}) 
+ v F({\bf x}) n({\bf p},{\bf x})
\nonumber\\\fl
&&+
i\hbar\left(u+{v\over2}\right)\nabla N({\bf x})\cdot\nabla_p f({\bf p},{\bf x})
+i\hbar{v\over 2}\nabla F({\bf x})\cdot\nabla_p n({\bf p},{\bf x}).
\end{eqnarray}
\end{itemize}
\subsubsection{The measured signal}
The measured signal is the total number of atoms excited to the 2S level, that 
is
\begin{eqnarray}\Label{MS1}
S(\Omega) &=& 
\int d^3{\bf x}\langle \psi_2^\dagger({\bf x})\psi_2({\bf x})\rangle .
\end{eqnarray}
We can use an ansatz
\begin{eqnarray}\Label{MS2}
\psi_2({\bf x}) &\approx&
\int d^3{\bf y}\,r({\bf x},{\bf y})\psi_1({\bf y}) 
\end{eqnarray}
to express the linearity of the process, in the sense of (\ref{bg2},\ref{bg3}), 
and then we can write
\begin{eqnarray}\Label{MS3}
\bar f({\bf y},{\bf x})  &\equiv&
\langle\psi_1^\dagger({\bf y}) \psi_2({\bf x})\rangle
\approx
\int d^3{\bf y}'\,
r({\bf x},{\bf y}')\langle\psi_1^\dagger({\bf y}) \psi_1({\bf y}') \rangle  .
\end{eqnarray}
It is reasonable to approximate the correlation function on the right hand side 
of this equation by a locally thermal form
\begin{eqnarray}\Label{MS4}
\langle\psi_1^\dagger({\bf y}) \psi_1({\bf y}') \rangle&\approx
N\left({{\bf y}+{\bf y}'\over 2}\right)
\exp\left(-{({\bf y}-{\bf y}')^2\over 2L^2}\right),
\end{eqnarray}in which
\begin{eqnarray}\Label{MS5}
L &\equiv& \hbar/\sqrt{2\pi{mkT}}.
\end{eqnarray}
The length $ L$ is sufficiently short in a thermal situation to treat the 
correlation function as being essentially local in (\ref{MS2},\ref{MS3}), and 
thus write
\begin{eqnarray}\Label{MS6}
\bar f({\bf y},{\bf x}) &\approx&
(\sqrt{2\pi}L)^3N({\bf y})r({\bf x},{\bf y}),
\end{eqnarray}
and using this we can then similarly derive
\begin{eqnarray}\Label{MS7}
S(\Omega) &\approx& \int d^3{\bf x}\int d^3{\bf y}{|\bar f({\bf y},{\bf x})|^2
\over 
(\sqrt{2\pi}L)^3N({\bf y})}
\quad .
\end{eqnarray}

\subsection{The homogeneous case}
Suppose now that $ V({\bf x}) = 0$, and $ n({\bf p},{\bf x})\equiv  
n_{\rm hom}({\bf p})$ is independent of ${ \bf x}$.
 There is then a solution of (\ref{TH-d9}) with 
$  f({\bf p},{\bf x})\propto n({\bf p}) $, also independent of $ {\bf x}$
and the corresponding equation of motion can be integrated over $ {\bf p}$
to give
\begin{eqnarray}\Label{TH-d10}
i\hbar {\partial F\over\partial t } &=&
\{\hbar(\omega_2-\omega_1)-2(u-v)N\}F + gEN e^{-i\Omega t} 
\end{eqnarray}
Thus in this case we find that
\begin{eqnarray}\Label{TH-d1001}
\bar f({\bf y},{\bf x}) &\to  &F
\exp\left(-{({\bf y}-{\bf x})^2\over 2L^2}\right)
\end{eqnarray}
and 
\begin{eqnarray}\Label{TH-d1002}
S(\Omega) &\to & {|F|^2 V\over {2\sqrt{2}} N}
\end{eqnarray}
where $ V$ is the volume of the homogeneous system being observed.  We can 
easily solve the equations (\ref{TH-d10}) to get
\begin{eqnarray}\Label{TH-D1003}
S(\Omega) &=&  {N_{\rm tot}(gE)^2\over{2\sqrt{2}}\hbar^2}
{\sin^2\left\{\left(\Omega-\Delta\right)T/ 2
\right\}
\over\left\{\left(\Omega-\Delta\right)/2\right\}^2}\
\\ \Label{TH-D1004}
&\to & {N_{\rm tot}(gE)^2\over{2\sqrt{2}}\hbar^2} 2\pi T\, 
\delta(\Omega-\Delta).
\\ \Label{TH-D1005}
\mbox{where } \Delta &=& \omega_2-\omega_1-{2(u-v)N\over\hbar}
\end{eqnarray}
This confirms  the result of (\ref{bg6}) with $ g_2 = 2$.

%%%%%%%%%%%%%%%%%%%%%%%%%%%%%%%%%%%%%%%%%%%%%%%%%%%%%%%%%%%%%%%%%%%%%%%
%\begin{figure}
%\hskip 18mm{\includegraphics[width=11.2cm]{figS}}
%\caption{\label{figS} Excitation as a function of laser detuning.  In all 
%cases 
%the laser is turned on for a time $ T=5000\mu{\rm s}$.  The solid curves are 
%solutions
%of the equations (\ref{TH-d9}), the dashed curves omit the ``streaming terms''
%on the second line of (\ref{bg6}).  a) $\omega_r = 3.9{\rm kHz}$ ; b)
%$ \omega_r = 780{\rm Hz}$. }

%\end{figure}

\FIGURE{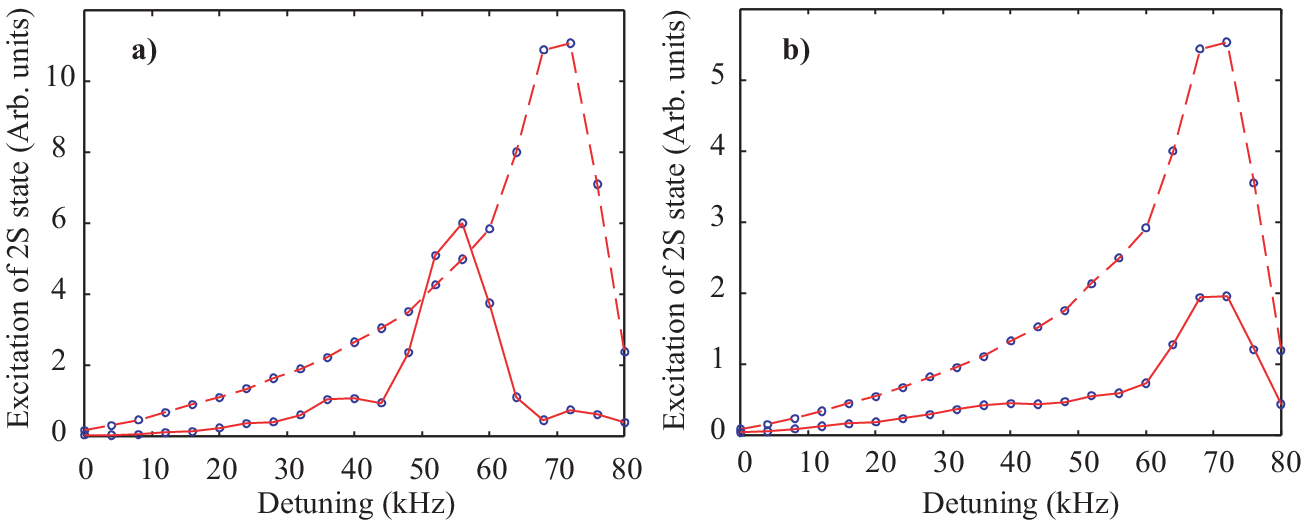}{Excitation as a function of laser detuning.  In all 
cases 
the laser is turned on for a time $ T=500\mu{\rm s}$.  The solid curves are 
solutions
of the equations (\ref{TH-d9}), the dashed curves omit the ``streaming terms''
on the second line of (\ref{TH-d9}).  a) $\omega_r = 3.9{\rm kHz}$ ; b)
$ \omega_r = 780{\rm Hz}$.}
 
%%%%%%%%%%%%%%%%%%%%%%%%%%%%%%%%%%%%%%%%%%%%%%%%%%%%%%%%%%%%%%%%%%%%%%

\subsection{One dimensional solutions of the equations of motion}
\label{onedsolutions}
The equations (\ref{TH-d9}) form a six-dimensional set, which are quite 
formidable to solve numerically, so to get an idea of their effect, we have 
done 
one-dimensional solutions, with trap frequency $ \omega_r$. The results are 
presented in Fig.\ref{figS} The solutions of 
the equations obtained by omitting the ``streaming term'' on the second line 
of 
(\ref{TH-d9})---corresponding to the result (\ref{bg6})---are given in the 
dashed curves.  It can be seen that for the trap frequency frequency of 
$ \omega_r=2\pi\times3.9{\rm kHz}$ , as used in the 
experiments, the excitation curve is very strongly affected, but for the 
frequency
$\omega'_r= \omega_r/5=2\pi\times780{\rm Hz}$, one 
fifth of that frequency, although there is a small quantitative difference in 
excitation, the shape is essentially unchanged. These should be compared with 
the excitation time $ T=500\mu{\rm s}$ for which $ \omega_rT/2\pi = 1.95 $, 
while $ \omega'_rT/2\pi = 0.39 $.
The conclusions we can draw from these one 
dimensional simulations are limited to the observation that the peak of the 
solid curve in Fig.\ref{figS} is shifted from the position which would be 
expected if one used the homogeneous system formulation of Sect.
\ref{background}.  The shift of the peak is such that one would see an apparent 
``factor of about $1.57$'', rather than the ``factor of 2'', and this would 
indicate that the true 1S--2S scattering length should be larger than the 
experiments have actually measured.  We are reluctant to treat the value 1.57 
as 
anything other than a qualitative result, but it seems that this would suggest 
that it might be wiser to use the theoretical value (\ref{a10201a}) in
analysing the condensate data rather than the apparent measured value 
(\ref{a10201b}).

\section{Cold collision frequency shift for an inhomogeneous condensed 
or partially condensed system}
The treatment in the presence of a condensate takes a completely different form 
from the full thermalized case. As before, the equations 
cannot be solved exactly, but approximate equations of motion which 
incorporate 
the features of the quasiparticle spectrum, as investigated in 
\cite{Gardiner0009371}, can be developed as follows.

\subsection{Approximate equation of motion}
The equations of motion and assumptions used in Sect.\ref{eqsmotion} are valid 
in this case as well, but the Wigner function methodology is not appropriate, 
since its use depends on hartree-Fock factorization.  

\subsubsection{Quasiparticle expansions}
As before, we suppose 
that the occupation of the 1S level is always very much 
greater than that of the 2S level, so that
$ \psi_1({\bf x},t) $ can be regarded as a known time-dependent operator. 
However, in this case we represent the time-dependent solution for 
$ \psi_1({\bf x},t)$
by a quasiparticle expansion of the form
\begin{eqnarray}\Label{n9}
\psi_1({\bf x},t) &=&
a_0(t)\Bigg({\xi({\bf x})\over \sqrt{n_0}}
+{1\over\sqrt{n_0}}\sum_n\left\{\alpha_n(t) u_n({\bf x}) 
+ \alpha^\dagger_n(t) v^*_n({\bf x}) \right\}\Bigg),
\end{eqnarray}
where the time-dependences of the  destruction operators are
\begin{eqnarray}\Label{n10}
\dot a_0(t) &=& -i(\omega_1 +\mu/\hbar) a_0(t),
\\ \Label{n11}
\dot \alpha_n(t) &=& -i(\epsilon_n/\hbar) \alpha_n(t).
\end{eqnarray}
The  most accurate forms for the amplitudes $ u_n({\bf x})  $, $  v^*_n({\bf 
x})$, and  values 
of the chemical potential $ \mu$ and  quasiparticle energies 
$\epsilon_n $ would be those determined by  the gapless formalism of 
Morgan and coworkers \cite{MorganThesis,Morgan2000a}.  However, for 
simplicity, 
we will use only the Hartree-Fock-Bogoliubov (HFB) method in the Popov 
approximation, which, as Morgan has shown, is a much better approximation than 
the pure HFB method. Our methodology depends only on the existence of 
approximations of the form (\ref{n9}--\ref{n11}), and so can be adapted to any 
approximation scheme of this general kind.

The equation for $ \psi_2({\bf x},t)$ is now linear and inhomogeneous, with 
operator coefficients.  We can develop approximate solutions using an ansatz 
of the form
\begin{eqnarray}\Label{n12}
\psi_2({\bf x},t)& =&
\quad{ a_0(t)\over\sqrt{n_0}}\Bigg({s({\bf x},t)}
+\sum_n\left\{\alpha_n(t) p_n({\bf x},t) 
+ \alpha^\dagger_n(t) q^*_n({\bf x},t) \right\}\Bigg)
\nonumber \\&&
 \qquad+\mbox{ terms involving } \psi_2({\bf x},0),
\end{eqnarray}
where the amplitudes $ s({\bf x},t)$, $ p_n({\bf x},t)$ and $ q_n({\bf x},t)$
are c-number amplitudes whose equations of motion are to be determined.

This form expresses in a more general sense the concept 
summarized in (\ref{bg2},\ref{bg3}),
that the driving field {\em coherently} rotates a small proportion of the the 
field $ \psi_1({\bf x},t)$ {\em linearly} into the 2S subspace.
Here we keep the linearity in the sense that
 (\ref{n12}) is a linear combination of components $ a_0(t)$,
$ a_0(t)\alpha_n(t)$, $\alpha_n^\dagger a_0(t)$ in terms of which the field 
$ \psi_1({\bf x},t)$ can be expressed. However, it can be seen that the 
evolution according to 
(\ref{TH-d2}) depends nonlinearly on the 1S field, and this leads in the 
approximation we are using to a dependence on nonoperator nonlinear functions 
of the field $ \psi_1$, in much the same way as the expression (\ref{n9}) gives 
linearized excitations of the condensate, whose equations of motion 
nevertheless 
depend on certain  averages of nonlinear functions of the field operator.
\subsubsection{Equations of motion}
From (\ref{n12}) we can express the coefficients in terms of thermal averages
\begin{eqnarray}\Label{n13}
\langle a_0^\dagger(t)\psi_2({\bf x},t)\rangle &=& \sqrt{n_0}\, s({\bf x},t),
\\ \Label{n14}
\langle a_0^\dagger(t)\alpha^\dagger_n(t)\psi_2({\bf x},t)\rangle &=&
 \sqrt{n_0}\,\bar N_n p_n({\bf x},t),
\\  \Label{n15}
\langle a_0^\dagger(t)\alpha_n(t)\psi_2({\bf x},t)\rangle &=&
 \sqrt{n_0}\,(\bar N_n +1) q_n^*({\bf x},t).
\end{eqnarray}
Here we assume the condensate-vapour system in the 1S state is in a thermal 
state, so that
\begin{eqnarray}\Label{n16}
\langle\alpha_n^\dagger(t)\alpha_n(t)\rangle &=& 
\bar N_n \equiv {1\over e^{\epsilon_n/kT}-1}.
\end{eqnarray}
Using these, it is straightforward to derive the set of coupled equations of 
motion:
%%%%%%%%%%%%%%%%%%%  END NEW  !!!
\begin{eqnarray}\fl\Label{e19}
i\hbar\dot s &=&
\left\{\hbar(\omega_2-\omega_1)-\mu+H +v[|\xi |^2+\bar n]\right\}s
+gE\xi e^{-i\Omega t},
\nonumber \\ \fl 
&&+v\xi \sum_m\Big\{\bar N_m[u^*_m + v^*_m]p_m
(\bar N_m+1)[v_m + u_m]q_m^*\Big\}
%%%  second equation
\\ \fl \Label{e20}
i\hbar \dot p_n &=&
\left\{\hbar(\omega_2-\omega_1)-\mu - \epsilon_n+H
 +v[|\xi |^2+\bar n]\right\}p_n 
+gEu_ne^{-i\Omega t},
\nonumber \\ \fl 
&& +\left\{ v[\xi ^*u_n +\xi v_n]\right\}s
\nonumber \\ \fl 
&& +v\sum_m\Big\{ \bar N_m[u_nu_m^*+v_m^*v_n]p_m
+(\bar N_m +1)[u_nv_m + u_mv_n]q_m^*\Big\}
%%% Third equation
\\ \fl \Label{e21}
i\hbar \dot q^*_n &=&
\left\{\hbar(\omega_2-\omega_1)-\mu + \epsilon_n+H
 +v[|\xi |^2+\bar n]\right\}q^*_n 
+gEv^*_ne^{-i\Omega t}
\nonumber \\ \fl 
&&+\left\{v[\xi u^*_n +\xi ^*v^*_n]\right
\}s
\nonumber \\ \fl 
&& +v\sum_m\Big\{ \bar N_m[v_n^*u_m^*+v_m^*u^*_n]p_m
+(\bar N_m +1)[v^*_nv_m + u_mu^*_n]q^*_m({
\bf x},t)\Big\}
.
\end{eqnarray}
Here $ \bar n({\bf x}) $ is the mean noncondensate density
\begin{eqnarray}\Label{e2101}
\bar n({\bf x})&=& \sum_n\left\{\bar N_n|u_n({\bf x})|^2
+(\bar N_n+1)|v_n({\bf x})|^2\right\}.
\end{eqnarray}
In terms of these amplitudes, the occupation of level 2 is
\begin{eqnarray}\fl\Label{e22}
\langle\psi^\dagger_2({\bf x},t)\psi_2({\bf x},t)\rangle
&=& |s({\bf x},t)|^2 + 
\sum_n\left\{\bar N_n|p_n({\bf x})|^2 + (\bar N_n +1)|q_n({\bf x},t)|^2\right
\}.
\end{eqnarray}
%\narrowtext 

%%%%%%%%%%%%%%%%%%%%%%%%%%%%%%%%%%%%%%%%%%%%%%%%%%%%%%%%%%%%%%%%%%%%%%%%%%%%%%%

\section{Simplified solutions}
The equations (\ref{e19}--\ref{e21}) are somewhat opaque, so in this section we 
will develop some simplifications so as to see the general structure of their 
predictions. The simplifications we shall make are quite drastic, and thus the 
results of this section may have only a qualitative significance. However, the 
full equations (\ref{e19}--\ref{e21}) should be rather accurate, and are 
by no means intractable, so that we will always be able to check the 
approximations explicitly.
\subsection{Development of equations}
In a previous paper \cite{Gardiner0009371} we developed approximate 
wavefunctions for the low lying quasiparticle modes, which we expect to 
dominate that part of the noncondensate fraction which exists in the same 
location as the condensate.  It was shown in this paper that the first 10 
eigenfunctions give nearly all the contribution to this part of the 
noncondensate fraction.
In this case for such low lying eigenfunctions, we can make some 
simplifications (which one would expect to be generally valid for any
kind of quasiparticle formalism):
\begin{itemize}
\item[i)] The occupations $ \bar N_n $ are very large, and we can neglect the
difference between $ \bar N_n $ and $ \bar N_n +1$.
\item[ii)] The eigenfunctions can be chosen real.
\item[iii)] There is little difference between the two types of 
eigenfunctions, in the sense that
$ v_n({\bf x}) \approx -u_n({\bf x})$.
\item[iv)] The quasiparticle energies $ \epsilon_n$ for these eigenfunctions 
are negligible  compared to the other mean field effects.
\end{itemize}
Putting in these simplifications, the equations for $ p_n$ and $ -q_n$ become 
the same as each other, so we can write $ q_n(t) \approx -p_n(t)$,  and the 
equations (\ref{e19}--\ref{e21}) simplify to
\begin{eqnarray}\Label{f1}
i\hbar\dot s &=& \left\{\hbar(\omega_2-\omega_1) -\mu + H({\bf x})
             +v[|\xi({\bf x})|^2 +\bar n({\bf x})]\right\} s 
\nonumber \\
&& + gE({\bf x})\xi({\bf x})e^{-i\Omega t}
\\ \Label{f2}
i\hbar \dot p_n &=& 
\Big\{\hbar(\omega_2-\omega_1) -\mu + H({\bf x})
 +v[|\xi({\bf x})|^2 +\bar n({\bf x})]\Big\}p_n
\nonumber \\
&&+ 4v\sum_m \bar N_m u_n({\bf x})u_m({\bf x})p_m
+ gE({\bf x})u_n({\bf x})e^{-i\Omega t}
.
\end{eqnarray}

\subsubsection{Condensate contribution}
The first of these equations, representing the condensate, is now independent 
of the others.  The method of solution is very similar to that used by 
Killian \cite{KillianThesis}, as follows.
We replace $ \mu$ using the Popov and Thomas-Fermi approximations in a 
\HFB\ theory, from which it follows that the condensate wavefunction 
approximately satisfies
\begin{eqnarray}\Label{f20101}
u|\xi({\bf x})|^2&=& 
\mu-V({\bf x})- 2u\bar n({\bf x})\,\mbox{ where }\,
\mu >V({\bf x})+ 2u\bar n({\bf x}) ;
\\ \nonumber &=& 0 \,\mbox{ otherwise.}
\end{eqnarray}
We can then eliminate $ \mu$ in the region where the condensate wavefunction is 
nonzero, which is where our formalism is valid and is also where the signal of 
interest  comes from. (Outside that 
region we can make no elimination, so there is an effective potential as in  
Fig.\ref{fig1}---the very tight attraction at 
the centre arises because $v$ is negative, and about 20 times larger 
than $u$.)

\FIGURE{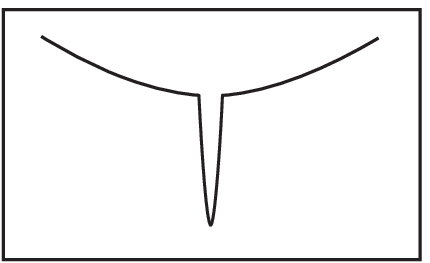}{Form of the effective potential.}
%\begin{figure}

%\includegraphics{fig1}
%\epsfig{file=fig1,width=6cm}

%\caption{\label{fig1}Form of the effective potential.}
%\end{figure}
Hence we 
expand in terms of the orthonormal eigenfunctions $ g(\omega,i,{\bf x})$ of the 
operator
\begin{eqnarray}\Label{f3}
{\cal H}_C 
&=& -{\hbar^2\nabla^2\over 2m}
             +(v-u)|\xi({\bf x})|^2 +(v-2u)\bar n({\bf x}).
\end{eqnarray}
Here $ \hbar\omega$ is the eigenvalue of $ {\cal H}_C$, and $ i$ represents all 
the other 
eigenvalues necessary to specify the state. Thus, we write
\begin{eqnarray}\Label{f4}
s({\bf x},t) &=& \sum_{\omega,i}\tilde s_{\omega,i}(t) g(\omega,i,{\bf x})
\\ \Label{f5}
gE({\bf x})\xi({\bf x}) &=& \sum_{\omega,i} I_{\omega,i} g(\omega,i,{\bf x}).
\end{eqnarray}
The solution for $ \tilde s_{\omega,i}(t)$ is then clearly
\begin{eqnarray}\Label{f6}
\tilde s_{\omega,i}(t) &=&  I_{\omega,i}e^{-i\Omega t}
\left(e^{-i(\omega_2-\omega_1+\omega-\Omega)t} -1\over
\hbar(\omega_2-\omega_1+\omega-\Omega)
\right).
\end{eqnarray}
The signal from the experiment is proportional to the total occupation, i.e.,
\begin{eqnarray}\Label{f7}
\int d^3{\bf x}|s({\bf x},t)|^2 &=&
\sum_{\omega,i}| I_{\omega,i}|^2
{\sin^2[(\omega_2-\omega_1+\omega-\Omega)t/2]\over
[\hbar(\omega_2-\omega_1+\omega-\Omega)/2]^2}
\\ \Label{f8}
&\to &{2\pi t \over \hbar^2}\sum_{\omega,i}| I_{\omega,i}|^2
\delta(\omega_2-\omega_1+\omega-\Omega).
\end{eqnarray}
The quantity $  I_{\omega,i}$ is given by
\begin{eqnarray}\Label{f9}
 I_{\omega,i} &=&g\int d^3{\bf x}\, g^*(\omega,i,{\bf x})\xi({\bf x})E({\bf x}) 
.
\end{eqnarray}
At this stage it becomes important to distinguish between the Doppler free 
case, where $ E({\bf x}) = {\cal E}$, a constant, and the Doppler sensitive 
case where $ E({\bf x})$ has a very rapid oscillation at the wavelength of the 
light being used, which makes a simple approximation scheme difficult.  We do 
not treat the Doppler-sensitive case in this paper. 
\subsection{Doppler free excitation}
In this case we write
\begin{eqnarray}\Label{f901}
 I_{\omega,i} &=&g{\cal E}\int d^3{\bf x}\, g^*(\omega,i,{\bf x})\xi({\bf x}) 
.
\end{eqnarray}
and  follow the reasoning of Killian \cite{KillianThesis}, that 
for relatively high energy excitations such as occur here, the integral of the 
product of a relatively smooth function, such as $\xi({\bf x}) $ and the  
eigenfunction $  g^*(\omega,i,{\bf x}) $ is dominated by the contribution
at the boundary of the classically allowed region, that is from 
$ {\bf x}$ such that
\begin{eqnarray}\Label{f10}
\hbar\omega =  (v-u)|\xi({\bf x})|^2 +(v-2u)\bar n({\bf x}).
\end{eqnarray}
Qualitatively, this occurs because the eigenfunction $  g(\omega,i,{\bf x})$ 
has a behaviour like that of an Airy function near the place where the energy 
eigenvalue is equal to the potential term in $ {\cal H}_C$, that is, where 
$ \omega$ satisfies (\ref{f10}). The Airy function and its integral are plotted 
in Fig.\ref{fig2}. Combining (\ref{f8}) and (\ref{f10}), we find that the 
main contribution comes from positions where
\begin{eqnarray}\Label{f101}
\Omega_C({\bf x}) &=& \omega_2-\omega_1 
+ {(v-u)|\xi({\bf x})|^2 +(v-2u)\bar n({\bf x})\over\hbar}.
\end{eqnarray}
\subsubsection{Case of a pure condensate}
In this case we set $ \bar n({\bf x}) =0$, and our results are the same as 
those 
of Killian \cite{KillianThesis,Killian2000a} for a pure condensate.

\FIGURE{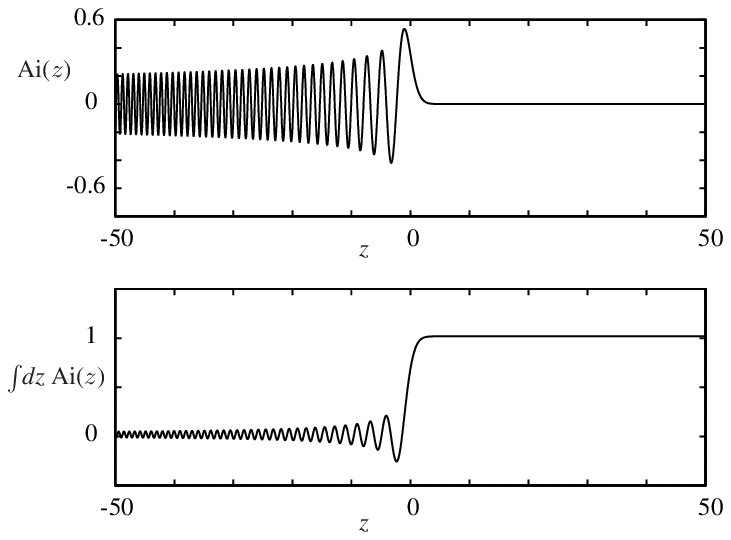}{Plot of the Airy function and its integral}

\subsubsection{Quasiparticle contribution}
The remaining equations can be further simplified by noting that because $ |v|
\gg |u|$, the 
spatial variation of $ p_n({\bf x})$ will be very much faster than that of 
$ u_n({\bf x})$.  Using this, we can make the ansatz
\begin{eqnarray}\Label{f11}
p_n({\bf x}) &=& u_n({\bf x})\bar P({\bf x})
\end{eqnarray}
and because of the slow variation of $ u_n({\bf x}) $ we can approximate
\begin{eqnarray}\Label{f12}
\nabla^2p_n({\bf x}) &\approx& u_n({\bf x})\nabla^2\bar P({\bf x}).
\end{eqnarray}
We use this ansatz, and the fact that 
when $ 1$ is neglected in comparison to $ \bar N_m$
\begin{eqnarray}\Label{f13}
\bar n({\bf x}) &\approx& 2\sum_m \bar N_m u_m({\bf x})u_m({\bf x}),
\end{eqnarray}
to conclude that, for any $ n$, 
the equations (\ref{f2}) become
\begin{eqnarray}\fl\Label{f14}
i\hbar{\partial p_n({\bf x}) \over\partial t}
\approx
\Big\{\hbar(\omega_2-\omega_1) -\mu + H({\bf x})    
 +v[|\xi({\bf x})|^2 +3\bar n({\bf x})]\Big\}p_n
+ g{\cal E}u_n({\bf x})e^{-i\Omega t}
.
\end{eqnarray}

\FIGURE{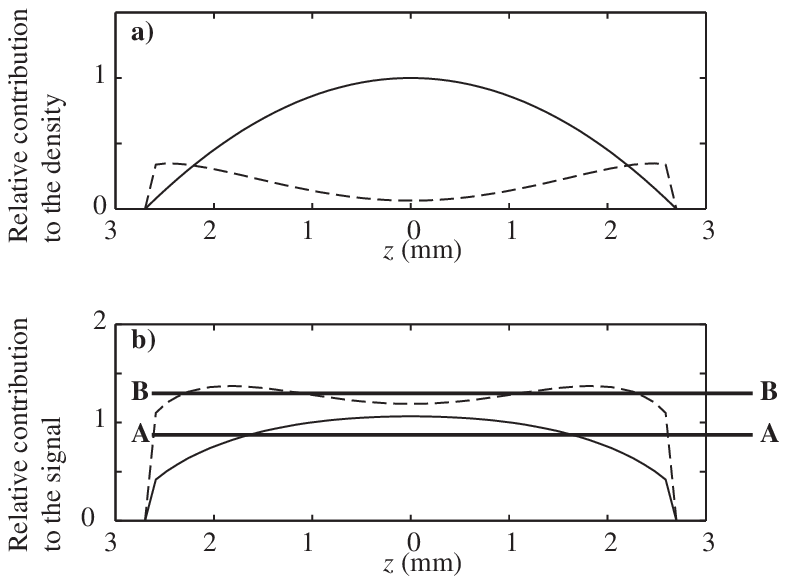}{a) Relative condensate (solid line) and 
quasiparticle (dashed line) contributions to the density;
b) Detunings generated by condensate ($ \Omega_{C}$, solid line) and 
quasiparticles ($ \Omega_{QP}$, dashed line); see text Sect.{\ref{sigsect}}.}

\noindent
The fact that we get the same equation for all $ n$ means that 
there are solutions of this equation of the form (\ref{f11})---that is,
our requirement that $ P({\bf x})$ be independent of $ n$ is verified.
As in the case for  equation for the condensate contribution, we can eliminate 
the chemical potential using
(\ref{f20101}), define the operator for the quasiparticles
\begin{eqnarray}\Label{f15}
{\cal H}_{Q}&\equiv&
 -{\hbar^2\nabla^2\over 2m}
             +(v-u)|\xi({\bf x})|^2 +(3v-2u)\bar n({\bf x}),
\end{eqnarray}
and expand in terms of its eigenfunctions $ f(\omega,i,{\bf x})$, using
\begin{eqnarray}\Label{f16}
p_n({\bf x},t) &=& \sum_{\omega,i}\tilde p_{n,\omega,i}(t) f(\omega,i,{\bf x})
\\ \Label{f17}
g{\cal E}u_n({\bf x}) &=& \sum_{\omega,i} J_{n,\omega,i} f(\omega,i,{\bf x}).
\end{eqnarray}
We proceed as for the condensate, and find that the quasiparticle signal 
is
\begin{eqnarray}\fl\Label{f18}
\sum_n\int d^3{\bf x}\,2\bar N_n |p_n({\bf x})|^2 &=&
\sum_{n,\omega,i}2\bar N_n|J_{n,\omega,i}|^2
{\sin^2[(\omega_2-\omega_1-\omega-\Omega)t/2]\over
[\hbar(\omega_2-\omega_1+\omega-\Omega)/2]^2}
\fl\\ \Label{f19}
&\to  &{2\pi t \over \hbar^2}\sum_{\omega,i}2\bar N_n|J_{n,\omega,i}|^2
\delta(\omega_2-\omega_1+\omega-\Omega),
\end{eqnarray}
where 
\begin{eqnarray}\Label{f20}
 J_{n,\omega,i} &=&g {\cal E} \int d^3{\bf x} f^*(\omega,i,{\bf x})
   	   	   	 u_n({\bf x}).
\end{eqnarray}
In this case,
the functions $u_n({\bf x}) $ are also very smooth compared to the 
eigenfunctions, and the integrals will be dominated by contributions from 
$ {\bf x}$ such that
\begin{eqnarray}\Label{f21}
\Omega_{QP}({\bf x})&=&\omega_2-\omega_1+  {(v-u)|\xi({\bf x})|^2 
+(3v-2u)\bar n({\bf x})\over\hbar}.
\end{eqnarray}

\section{Experimental Comparison}\label{sigsect}
The equations (\ref{f10},\ref{f21}) can be viewed as the fundamental equations 
for determining the position in the system from which the measured signal 
(i.e., the total occupation of the 2S state) originates, although they are 
definitely only approximations to the equations (\ref{e19}--\ref{e21}). It is 
apparent the 
the resonance condition for the condensate component, (\ref{f10}) is 
significantly different from that of the quasiparticle component, (\ref{f21}), 
especially as $ v \approx -22 u$.

\FIGURE{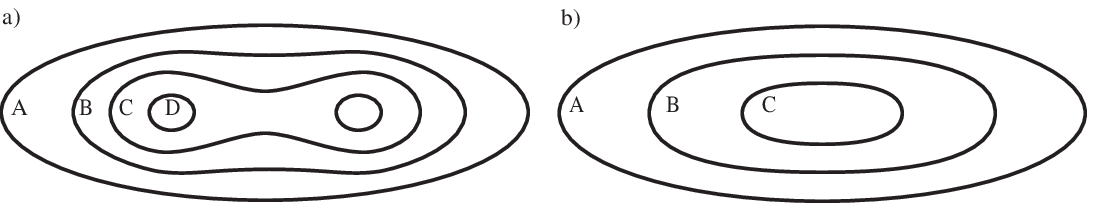}{Illustration of the contours of excitation 
expected for a) The quasiparticle component and b) The condensate 
component.  The letters A--D correspond to the frequencies
$\Omega_{\rm A}$--$\Omega_{\rm D}$.  Note that the highest frequency
$\Omega_{\rm D}$ gives no condensate signal.}

In order to evaluate the full signal from the two components, it will be 
necessary to evaluate the sums (\ref{f8},\ref{f19}) in the case that the 
eigenfunctions are almost Airy functions, and this is not a trivial task.  In 
the case that we can approximate the effective potentials by a harmonic 
approximation, we can show, as in \cite{Killian2000a,KillianThesis}, that we 
can take the intuitive form
\begin{eqnarray}\Label{sig1}
 S(\Omega) &\propto& \int d^3{\bf x}\Big[|\xi({\bf x})|^2
\delta\Big(\Omega-\Omega_{C}({\bf x})\Big) +\bar n({\bf x})
\delta\Big( \Omega-\Omega_{QP}({\bf x})\Big)\Big],
\end{eqnarray}
in which $\Omega_{C} $ and $ \Omega_{QP}$ are given by (\ref{f101}) and 
(\ref{f21})

The measured signal is seen form (\ref{sig1}) indeed to be the result 
of an inhomogeneous summation---at a given $\Omega$ the condensate signal 
(first 
term) and the quasiparticle signal (second term) come from different regions of 
the cloud.  Using the eigenfunctions from \cite{Gardiner0009371} we show in 
Fig.
\ref{QuasiSignal} how this happens.  Along the line A--A, corresponding to $ 
\Omega=\Omega_{\rm A}$ the two signals come from very different positions, 
while at 
$ \Omega=\Omega_{\rm B}$ there is no condensate signal, and the quasiparticle 
signal is in fact itself derived from multiple positions.  What can be seen is 
that the signal with highest detuning is in fact derived from the quasiparticle 
component of the signal.

The result of this dual source of the excitation signal is to produce a 
spectrum which has two very distinctive components, as shown in Fig.\ref{fig6},
where the condensate component is qualitatively the same as one would expect 
from a Thomas-Fermi treatment, and the quasiparticle component is dominated by 
a 
large detuning component, which forms a distinct ``foot'', protruding from the 
from the large detuning end of the spectrum.  At a given temperature, this 
feature is more distinct at lower condensate numbers.

The agreement with the experimental data of \cite{Fried1998a} is rather good.  
We have taken the total output signal in the form (\ref{sig1}), and used
the \emph{calculated} value, (\ref{a10201a}), of the 1S--2S scattering length, 
since the results of Sect.\ref{onedsolutions} indicate that the apparent value 
resulting from a measurement 
in a noncondensed system using the simple $ g_2=2$ method appropriate to a 
homogeneous system is likely to be significantly lower than the correct value.

\section{Conclusion}
We believe that in this paper we have resolved the apparent anomaly, that the 
hydrogen condensate appeared to be incoherent, and therefore have 
$ g_2=2$, rather than $ g_2=1$, as expected for a true condensate.  There are 
two  reasons for the appearance of this anomaly:
\begin{itemize}
\item [i)] The use of a formalism only strictly valid for the case of a 
homogeneous untrapped condensate is not justified with the trap geometry 
employed for the experiment.  The time during which the system is illuminated 
by the exciting illumination must always be significantly shorter than
the shortest trap period for this to be valid.

\item[ii)]
The more careful treatment of quasiparticle effects leads to higher frequency 
shifts than would be expected from a treatment based on the assumption that the 
condensate is pure.
\end{itemize}
We have also established that the presence of quasiparticles causes a {\em 
splitting} in the resonance, rather than a simple shift, and this leads to a 
a very characteristic ``foot'' in the excitation spectrum, of which there is 
already some experimental evidence.

\FIGURE{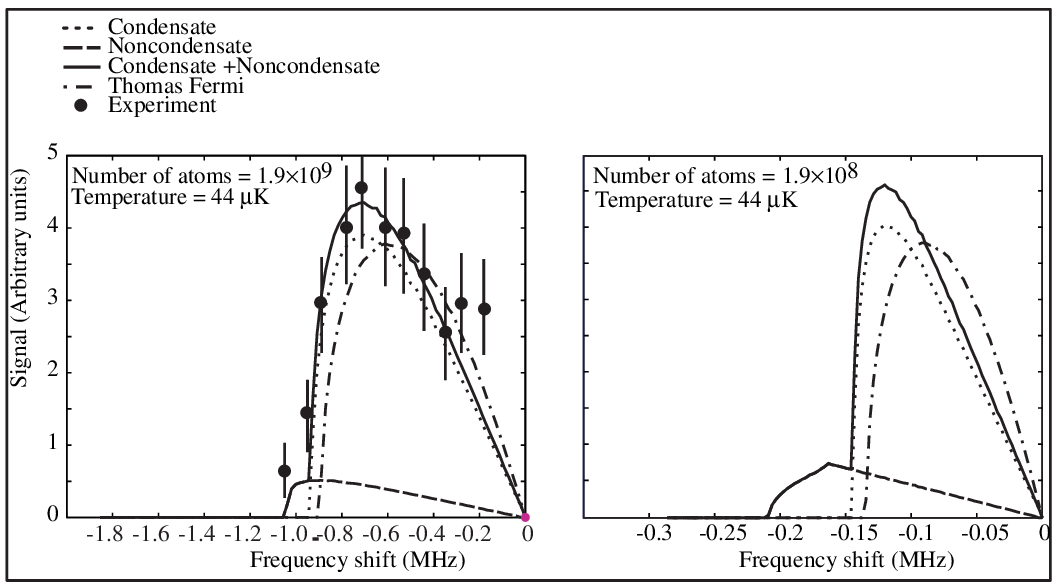}{Predicted and measured excitation spectra for
a) Parameter regime of the experiments \cite{Fried1998a,Killian1998a};
b) One tenth of the number of condensate atoms.
Notice that the ``foot'' is much more prominent in the latter. }

Finally, it must be emphasized that our treatment still makes many 
approximations.  In future work we will include mean-field effects in 
out computations of eigenfunctions, and shall also develop methods of 
solving the full equations so derived, including the  Doppler 
sensitive  case, which has not been amenable to a simple approximate 
treatment.

\ack
We wish to thank Dan Kleppner, Tom Greytak, Stephen Moss, Lorenz Willman and 
Kendra Vant for useful discussions about the hydrogen condensate experiments 
and hospitality in {MIT}.  This research was supported by the Marsden Fund of 
the Royal Society of new Zealand under contract PVT-902.

\section*{References}
\bibliography{HydSpec}

\end{document}